# Temperature selective thermometry with sub-microsecond time resolution using dressed-spin states in diamond

*Jiwon Yun‡, Kiho Kim‡, Sungjoon Park and Dohun Kim\**

*Department of Physics and Astronomy, and Institute of Applied Physics, Seoul National University, Seoul 08826, Korea*

‡ *These authors contributed equally to this work.*
*\*Corresponding author: dohunkim@snu.ac.kr*

**ABSTRACT**

Versatile nanoscale sensors that are susceptible to changes in a variety of physical quantities often exhibit limited selectivity. This paper reports a novel scheme based on microwave-dressed spin states for optically probed nanoscale temperature detection using diamond quantum sensors, which provides selective sensitivity to temperature changes. By combining this scheme with a continuous pump-probe scheme using ensemble nitrogen-vacancy centers in nanodiamonds, a sub-microsecond temporal resolution with thermal sensitivity of 3.7 $\mathrm{K\cdot Hz^{-1/2}}$ that is insensitive to variations in external magnetic fields on the order of 2 G is demonstrated. The presented results are favorable for the practical application of time-resolved nanoscale quantum sensing, where temperature imaging is required under fluctuating magnetic fields.

## 1. Introduction

Sensing local temperatures with the high spatiotemporal resolution is an important techniq ue that can be utilized in a wide range of fields, including electronics and biology.[1-4] Among recently developed time-resolved nanothermometers,[5-9] the nanodiamond embedding color centers such as nitrogen-vacancy (NV) center, silicon-vacancy (SiV) center, and germanium-vacancy (GeV) center have been highlighted as promising temperature sensors [10-20] that can be operated under ambient conditions and at room temperature based on their newly developed quantum sensing properties,[21, 22] together with the material's chemical inertness, high photostability, and biocompatibility.[23]

NV centers have spin-triplet ground states in which the magnetic quantum number $m_s = 0$ and $m_s = \pm 1$ sublevels are energetically separated by temperature $T$-dependent zero-field splitting ($D(T)$) due to crystal field.[24-26] The relative transition frequencies between sublevels are susceptible to changes in environmental factors such as magnetic fields,[27, 28] electric fields,[29, 30] strain,[31-34] and temperature[35, 36] meaning NV centers can serve as versatile nanoscale sensors. For primary sensing applications, many NV-center-based magnetometry techniques have been developed, including continuous wave optically detected magnetic resonance (cw-ODMR)-based methods,[37] pulsed ODMR-based methods,[38] and more advanced dynamical decoupling pulse-sequence-based magnetometry for improving spin coherence time and field sensitivity by reducing the interactions between NV centers and other adjacent spins, such as $^{13}$C nuclear spins.[39-41] The magnteometry techniques can also be applied to measuring nearby current flows by applying the magnetometry to an NV center ensemble under external magnetic field to measure the local vector magnetic field induced from the local current. Recently, measurements of current flows in nanoelectric devices such as graphene using the NV center as a probe were reported.[42]

Unlike magnetometry, NV-center-based thermometry utilizes energy level shifts related to direct changes in $D(T)$. Because the relative energy level difference is the only accessible information in an experiment, the careful calibration or experimental fixing of environmental factors other than the desired sensing quantity is necessary. This is particularly problematic in thermometry, where magnetic-field-induced sublevel shifts often overwhelm temperature-related changes.[35] Several studies have proposed approaches to suppressing magnetic field effects. For example, pulse sequences such as the thermal echo, thermal Carr-Pucell-Meiboom-Gill sequence, and D-Ramsey sequence have been developed for high-temperature sensitivity.[10, 13] Real-time monitoring methods have also been developed to reduce magnetic field effects.[14, 15, 17] Although these studies have shown merit with carefully pre-calibrated settings, robustnes s to a wide range of field magnitudes and directions has not been demonstrated and existing methods often exhibited limited temporal resolution because a long phase accumulation time per sequence is required.

The limited temporal resolution is an obstacle when thermal gradients of the nanodevices are the object of interest, since the thermal conductance in nanoelectric devices tend to be high in line with the high electrical conductance which is usually required for these devices. This makes it difficult to image the local thermal distribution of the device because of the fast heat dissipation. To resolve the local heat sources on the device before the thermal distribution reaches its steady state, a measurement technique with high temporal resolution is required. Therefore, for measurement of the local current and temperature using the same probe, it is important to develop continuous measurement-based thermometry sequences that can be applied in non-zero external magnetic field environments.

Here, we propose and demonstrate an NV-center-based micron-scale time-resolved thermometry method using microwave-dressed spin states, which is applicable to even non-zero external magnetic field environments. In a dressed spin space,[43-45] temperature-induced frequency shifts can be detected using net spin-zero eigenstates. Temperature information can be obtained by measuring up to six frequency components within the ODMR curve, making the proposed method applicable to real-time or time-resolved measurements. We demonstrate time-resolved temperature measurements with 10-μm spatial resolution and 50 ns temporal resolution using NV centers in nanodiamonds with natural tolerance to the varying magnitudes and directions of external magnetic fields.[13] When combined with well-developed independent magnetometry methods, the proposed method provides an effective route for a wide range of quantum sensing and chip-scale imaging applications for which selectivity and sensitivity to sensing quantities are both important.

**Figure 1**a presents a schematic of the proposed quantum sensing setup, where the NV centers in diamonds are subjected to microwave power-induced Joule healing of the coplanar waveguide (CPW), as well as an external magnetic field of arbitrary magnitude and direction. By ignoring the effects of the nitrogen or $^{13}$C nuclear spins in diamond, the Hamiltonian can be described as (we adopt $h$=1)

$$H_0 = D(T)S_z^2 + \gamma_e \boldsymbol{B} \cdot \boldsymbol{S} + E(S_x^2 - S_y^2),  \tag{1}$$

where $\gamma_e = 2.8~\text{MHz} \cdot \text{G}^{-1}$ is the gyromagnetic ratio of the electron spin, $\boldsymbol{B}$ is the external magnetic field, $E$ is the off-axial strain, and $S_{i,~i=x,y,z}$ ($\boldsymbol{S}$) is the $i^{th}$ component (vector) of spin-1 matrices with their NV axis set as the $z$-direction. Since this work is focused on using continuous wave ODMR measurement, where microwave powers that are stronger than the energy level splitting from the hyperfine interaction induced by the nitrogen or $^{13}$C nuclear spins in diamond are used, and since the change of temperature only affects $D(T)$, we limit the discussion of the Hamiltonian to this simple form.

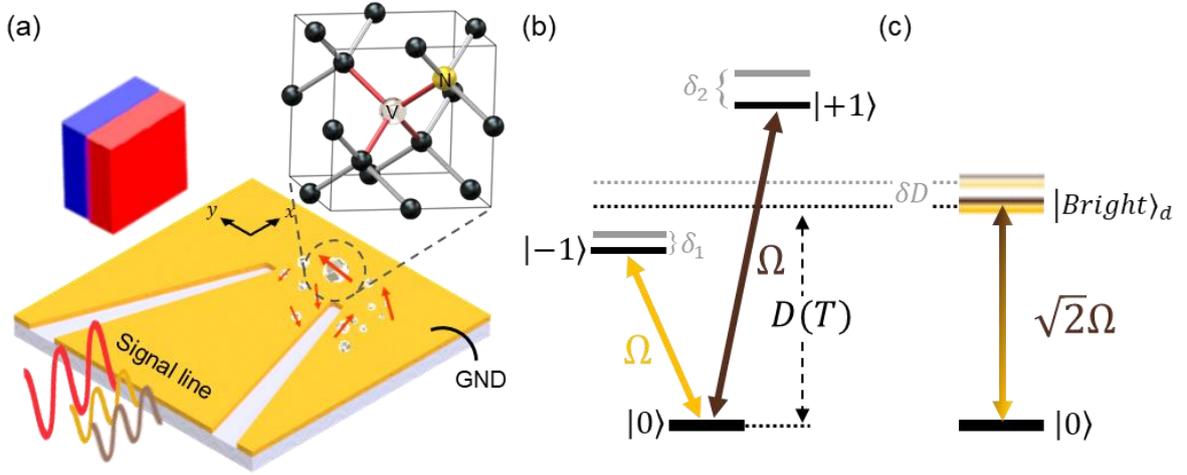

**Figure 1.** (a) Schematic of temperature measurement using NV centers in nanodiamonds under an arbitrary external magnetic field. The microwaves for spin state measurement and heating are fed into the shorted-end coplanar waveguide (CPW) signal line. (b) The energy level of the NV center ground state, where $D$ represents the zero-field splitting of the spin-triplet and $\Omega$ represents the Rabi oscillation frequency resonant to the $|0\rangle \leftrightarrow |\pm 1\rangle$ transition. (c) State transition between the $|Bright\rangle_d$ state in the dressed-state basis and $|0\rangle$ state when applying the microwave shown in (b).

As shown in Figure 1b, typical quantum sensing measures changes in the transition frequency between bare spin states ($|0\rangle$, $|-1\rangle$, and $|+1\rangle$) by applying corresponding

microwave pulses. By applying the double rotating frame approximation, the effects of small changes in $D(T)$ ($\delta D(T)$) and $\boldsymbol{B}$ ($\delta\boldsymbol{B} = (\delta B_x, 0, \delta B_z)$) lead to a linear change in the frequencies, where $\delta_{1(2)} = (\delta D(T) - (+)\gamma_e \delta B_z) + 3\frac{\gamma_e B_x}{D(T)}\gamma_e \delta B_x$. Therefore, extracting $D(T)$ information from the measurement of $\delta_{1(2)}$ requires prior knowledge regarding $\boldsymbol{B}$. A lack of this knowledge leads to significant temperature measurement errors based on the large sensitivity of NV centers to magnetic fields ($\gamma_e$ =2.8 MHz·G$^{-1}$ versus $dD/dT$ ~74 kHz·K$^{-1}$ [35]).

We address this issue by changing the eigenbasis to microwave-dressed spin states defined by $|0\rangle_d = |0\rangle$, $|Bright\rangle_d = (|+1\rangle + |-1\rangle)/\sqrt{2}$, and $|Dark\rangle_d = (|+1\rangle - |-1\rangle)/\sqrt{2}$, as shown in Figure 1c. In the dressed spin state basis, the eigenstates exhibit nearly zero Zeeman splitting, opening a pathway for the selective measurement of temperature. To achieve this basis, we apply two microwaves that are each resonant to the energy level splitting between $|0\rangle \leftrightarrow |+1\rangle$ and $|0\rangle \leftrightarrow |-1\rangle$ and shares the same Rabi frequency $\Omega$ to move into the double rotating frame of this Hamiltonian. To keep the double rotating frame valid, it is important to prepare the external magnetic field that induces the Zeeman splitting to be larger than the Rabi frequency of a single microwave ($\gamma_e B_z > \Omega$), since otherwise the two transitions cannot be independently addressed. In addition, when the external magnetic field satisfies the intermediate range of $E \ll \gamma_e |\boldsymbol{B}|$ and $\gamma_e B_x \ll D(T)$, the Hamiltonian in the dressed basis and in its double rotating frame becomes

$$H_{\text{tot,d}} = \begin{pmatrix} \delta_D & \sqrt{2}\Omega & \delta_B \\ \sqrt{2}\Omega & 0 & 0 \\ \delta_B & 0 & \delta_D \end{pmatrix}, \qquad (2)$$

where $\Omega$ is the Rabi frequency of a single microwave, $\delta_D = \frac{\delta_1 + \delta_2}{2} \approx \delta D(T) + 3\frac{\gamma_e B_x}{D(T)}\gamma_e \delta B_x \approx \delta D(T)$ is the response of the energy level of the $|Bright\rangle_d$ and $|Dark\rangle_d$ state, and the coupling between the $|Bright\rangle_d$ and $|Dark\rangle_d$ state is $\delta_B = \frac{\delta_2 - \delta_1}{2} \approx \gamma_e \delta B_z$. The transition Rabi frequency between $|0\rangle_d$ and $|Bright\rangle_d$ becomes $\sqrt{2}\Omega$ in this basis, which results as an increase in linewidth when the ODMR measurement is conducted. By directly driving $|0\rangle$ to $|Bright\rangle_d$, the influence of the fluctuation $\delta B_z$ is negligible in the strong driving regime $\delta_B \ll \sqrt{2}\Omega$ and the effects of the transverse field fluctuation $\delta B_x$ are reduced by a factor of $3\gamma_e B_x / D(T)$. Additionally, when the dressed spin state is used under an intermediate external magnetic field ($E \ll \gamma_e |\boldsymbol{B}|$), the energy levels become robust to changes in $E$. In Supporting Information S1, we describe this theoretical analysis in greater detail.

## 2. Results and Discussion

We first validated the robustness of the dressed state to changes in the external magnetic field at a constant temperature using a single NV center in a type-Ib bulk diamond (See Supporting Information S2 for details). For this experiment, we applied a variable external magnetic field in a fixed arbitrary direction. The upper panel of **Figure 2**a presents the ODMR obtained by sweeping a single microwave tone frequency. The two resonant peaks at $\omega_1$ and $\omega_2$ for the NV center represent energy level splitting between the $|0\rangle \leftrightarrow |-1\rangle$ and $|0\rangle \leftrightarrow |+1\rangle$ states. We then measured the dressed state cw-ODMR (DS-ODMR) spectrum by applying two microwaves simultaneously with equal frequency detuning from $\omega_1$ and $\omega_2$ to the NV center to control it within its double rotating frame, as shown in Figure 1a (bottom panel).

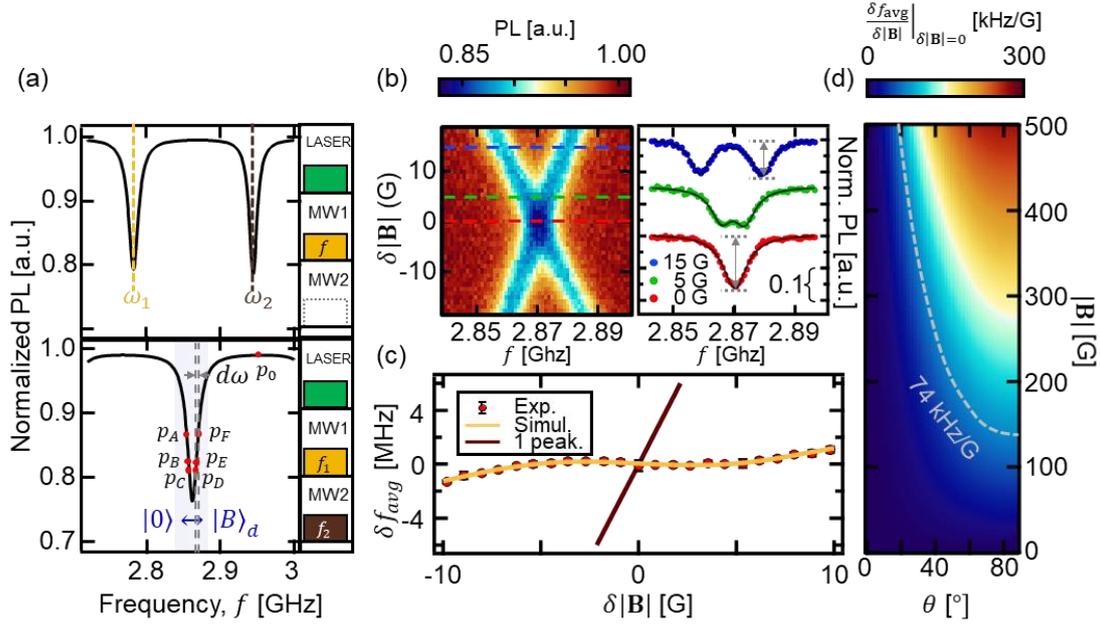

**Figure 2** (a) The upper part shows the example of an ODMR spectrum of a single NV center in a finite magnetic field obtained using the sequence on the right. The transition frequencies between $m_s = 0$ and $m_s = \pm 1$ are denoted as $\omega_1$ and $\omega_2$, respectively. The lower part shows the example of a DS-ODMR spectrum measured by applying two microwaves (MW1, MW2) with a frequency difference of $\omega_2 - \omega_1$ simultaneously, as shown in the right pulse sequence. The center dip represents the resonance peak formed by the bright state of the MW dressed state. The x-axis of the ODMR spectra represents the average frequency of the two microwaves used in the sequence. The red points represent the measurement frequencies used for thermometry, as described in the main text. (b) Left: 2D image of the DS-ODMR results while sweeping the external magnetic field. Right: Line cuts along a fixed magnetic field detuning. (c) Measured $f_{avg}$ changes at a fixed temperature depending on external magnetic field detuning. The red dots represent the change in $f_{avg}$ converted from fluorescence signals and the yellow line represents the simulated results for the DS-ODMR under the same conditions using QuTip. The brown line represents the converted frequency difference of the center frequency of a single peak in the single-microwave cw-ODMR. (d) Simulation results for the attenuated $\left.\frac{\delta f_{avg}}{\delta |\mathbf{B}|}\right|_{\delta |\mathbf{B}|=0}$ of the proposed method with changes in the strength and angle of the external magnetic field. The grey dotted line represents the line at which $\left.\frac{\delta f_{avg}}{\delta |\boldsymbol{B}|}\right|_{\delta|\boldsymbol{B}|=0} =$ 74 $\text{kHz} \cdot \text{G}^{-1}$. PL: Photoluminescence.

When sweeping the two microwave frequencies as $f_1 = f - (\omega_2 - \omega_1)/2$ and $f_2 = f + (\omega_2 - \omega_1)/2$ with a common center microwave frequency $f$, a Lorentzian curve is measured with the resonance at $f_{avg} = (\omega_1 + \omega_2)/2$ (double resonance with $f_1 = \omega_1, f_2 = \omega_2$), which represents the transition between $|0\rangle \leftrightarrow |Bright\rangle_d$ states in the dressed state basis, where the variation in $f_{avg}$ is directly proportional to $\delta_D$. The linewidth $\Gamma$ of this curve is determined by the Rabi frequency $\sqrt{2}\Omega$ coming from the two microwaves and the power of the green laser used for the continuous ODMR measurement.[38] The power of the green laser was limited to few tens of microwatts during all of the experiments, leaving the linewidth $\Gamma$ to be in the same order with the Rabi frequency $\sqrt{2}\Omega$. For consistency, from now on the term "center frequency" means the common center microwave frequency used in the DS-ODMR curve.

The robustness to external magnetic fields inherent in the Hamiltonian in **Equation 2** shows that measurement of the change in temperature can be conducted by measuring only a single center frequency point from the DS-ODMR curve. This property is useful when measuring a 2D image of the surface temperature, as a single measurement can be used as a single image. However, a single point measurement is vulnerable to environmental changes such as the photon counts of the NV center or the linewidth of the DS-ODMR curve. For practical measurement of the temperature, two methods, the three-point and six-point method, were used for the extraction of $\delta f_{avg}$ from this Lorentzian curve. Even though the advantage of measuring temperature directly from a single measurement vanishes from these methods, there are still advantages coming from using the dressed-state basis, which is that we can reduce the measurement time into half compared to applying this measurement method to the single-microwave cw-ODMR curve, which requires the frequency change measurement for both ODMR resonance peaks. For a simple three-point measurement (the six-point method is discussed below), $\delta f_{avg}$ can be extracted as

$$\delta f_{avg} = \frac{p_B - p_E}{p_B + p_E - 2p_0} \frac{\Gamma}{\sqrt{3}}, \tag{3}$$

where $\Gamma$ is the linewidth of the DS-ODMR Lorentzian curve, and $p_B$, $p_E$ and $p_0$ are the measured photon counts at the center microwave frequencies $f_B = f_{avg} - \Gamma/2\sqrt{3}$, $f_E = f_{avg} + \Gamma/2\sqrt{3}$, and the $f_0$ that corresponds to the background photon counts (see Figure 2a).[14] Using the three-point method, the effect of the temperature change was measured while securing robustness to the emitted photon counts from the NV center.

The middle panel in Figure 2b presents the DS-ODMR as a function of $\delta|B|$ with respect to a reference field $|B|$ of 47 G. As shown in the left and right panels of Figure 2b, the profile of the Lorentzian curve changes from a double Lorentzian peak profile when $\delta|B|$ is large to a single Lorentzian peak profile when $\delta|B|$ goes to 0. The increased contrast and linewidth of the Lorentzian curve at $\delta|B| = 0$ is a property showing that the curve is a result of the double microwave transition, as the transition Rabi frequency shifts from $\Omega$ to $\sqrt{2}\Omega$ as discussed above. When the $f_{avg}$ variation is extracted from Figure 2b using Equation 3, where the linewidth $\Gamma = 10.7$ MHz was used for the three-point method, the dressed state exhibits attenuation in $f_{avg}$ variation as a result of $\delta|B|$ down to approximately 50 kHz·G$^{-1}$, which is approximately 56 times smaller than 2.8 MHz·G$^{-1}$ of the bare spin states (Figure 2c, brown solid line), which is consistent with the prediction given by **Equation 2**. This result also agrees well with numerical simulations based on the QuTip package.[46, 47] As $\delta|B|$ increases, $\delta_B$ eventually becomes comparable to $\sqrt{2}\Omega$ and the $|0\rangle_d$ - $|Bright\rangle_d$ state subspace can no longer be considered to be decoupled from the $|Dark\rangle_d$ state, resulting in non-negligible deviations caused by $\delta|B|$. Figure 2d presents the numerical simulations (see Supporting Information S3 for details) for the magnetic-field-induced error $\left.\frac{\delta f_{avg}}{\delta|B|}\right|_{\delta|B|=0}$ as a function of the magnitude and orientation of **B**.

The gray dashed curve represents a contour of 74 kHz·G$^{-1}$ corresponding to 1 K·G$^{-1}$ when converted into temperature, demonstrating that the orientation and magnitude of the operating field for this method can vary widely while maintaining a sub-Kelvin magnetic field-induced temperature error.

We applied the dressed-state sensing method to ensemble NV centers to demonstrate time-resolved thermometry. **Figure 3**a presents dispersed nanodiamonds with an average diameter of 100 nm on top of a CPW on a cover glass substrate. From the single-tone ODMR curve with $|B|$ = 190 G, we selected the two peaks stemming from the NV ensembles whose axes are within approximately 29° with respect to the direction of $B$ to ensure temperature selectivity, as discussed above.[48] Temperature variation was induced by a 3-μs-long microwave pulse with a carrier frequency of 2.8 GHz, power of 48 dBm, and repetition rate of 100 kHz applied to the CPW. For the microwave used to measure the time-resolved temperature, we used the microwave power that is approximately 30dB weaker than the power used for heating to minimize the heating effect coming from this microwave. To compensate for the fluctuation in linewidth, average fluorescence, and long-term drift, we introduced the six-point measurement method (Figure 3b), which uses the measured photon counts from $p_A$ to $p_F$ at corresponding center frequencies $f_A$ to $f_F$, as shown in Figure 2a, where each frequency is chosen as follows.

$$f_A = f_{avg} - \Gamma/2\sqrt{3} - d\omega,$$

$$f_B = f_{avg} - \Gamma/2\sqrt{3},$$

$$f_C = f_{avg} - \Gamma/2\sqrt{3} + d\omega, \quad (4)$$

$$f_D = f_{avg} + \Gamma/2\sqrt{3} - d\omega,$$

$$f_E = f_{avg} + \Gamma/2\sqrt{3},$$

$$f_F = f_{avg} + \Gamma/2\sqrt{3} + d\omega,$$

where $d\omega$ = 1 MHz is a fixed value used and the linewidth of the DS-ODMR used in Fig. 3 was $\Gamma = 11.5$ MHz.

We used the same two main points that were used for the center frequency calculation in Figure 2 ($p_B$, $p_E$), but instead of obtaining the slope of the Lorentzian from a fixed $\Gamma$ value, we calculated the slope of the Lorentzian for each measurement period using the additional four points ($p_A$, $p_C$, $p_D$, $p_F$).

The temperature change can be extracted as (see Supporting Information S4 for details)

$$\delta T = \frac{p_B - p_E}{(p_A - p_C) - (p_D - p_F)} \frac{2d\omega}{dD/dT} \tag{5}$$

with a shot-noise-limited sensitivity of

$$\eta_T = 0.77 \frac{1}{dD/dT} \frac{\Gamma}{C_0} \frac{1}{\sqrt{R}}, \tag{6}$$

where $C_0$ is the luminescence contrast of the DS-ODMR curve and $R$ is the measured photon count per second (cps).

To combine the six-point measurement method and the pump-probe type time-resolved measurement, we used the frequency modulation mode of the signal generator to change the center frequency of the two microwaves applied to the NV center stepwise while continuously reading out the emitted fluorescence. A total period of 60 μs was used, with a 10 μs time range allocated for each center frequency, as shown in Figure 3c. The measurement was conducted using a time-tagged photon counting module, enabling the result to be time-resolved. During this measurement process, we used a 16 ns bin for the accumulation of the emitted photons. The frequency shifting of the DS-ODMR microwave, the pulse envelope for the heating microwave, and the time-tagged photon counting process all shared the same trigger which was applied at the start of the 60 μs period to synchronize the whole measurement sequence. During the continuous measurement, there exists some dead time due to the rise time of the signal generator (3.5 μs) when using the frequency shifting function. Figure 3d shows the fluorescence data where the data coming from different center frequencies were stacked based on the delay within the 10 μs time range, and with the dead time removed. For each delay time, we applied Equation 5 to the fluorescence $p_A$ to $p_F$, which converted the fluorescence data into temperature change, as shown in Figure 3e. To reduce the temperature noise, we box averaged 3 bins during the temperature conversion process, limiting our time resolution to 48 ns. When the heating microwave is turned on, a sudden drop of the fluorescence was observed, which is thought to come from a sudden torsion of the cover glass substrate. However, since the six-point method is robust to such sudden changes in fluorescence, contrast, and linewidth, the effect of the change of fluorescence was canceled out after converting the fluorescence data into temperature.

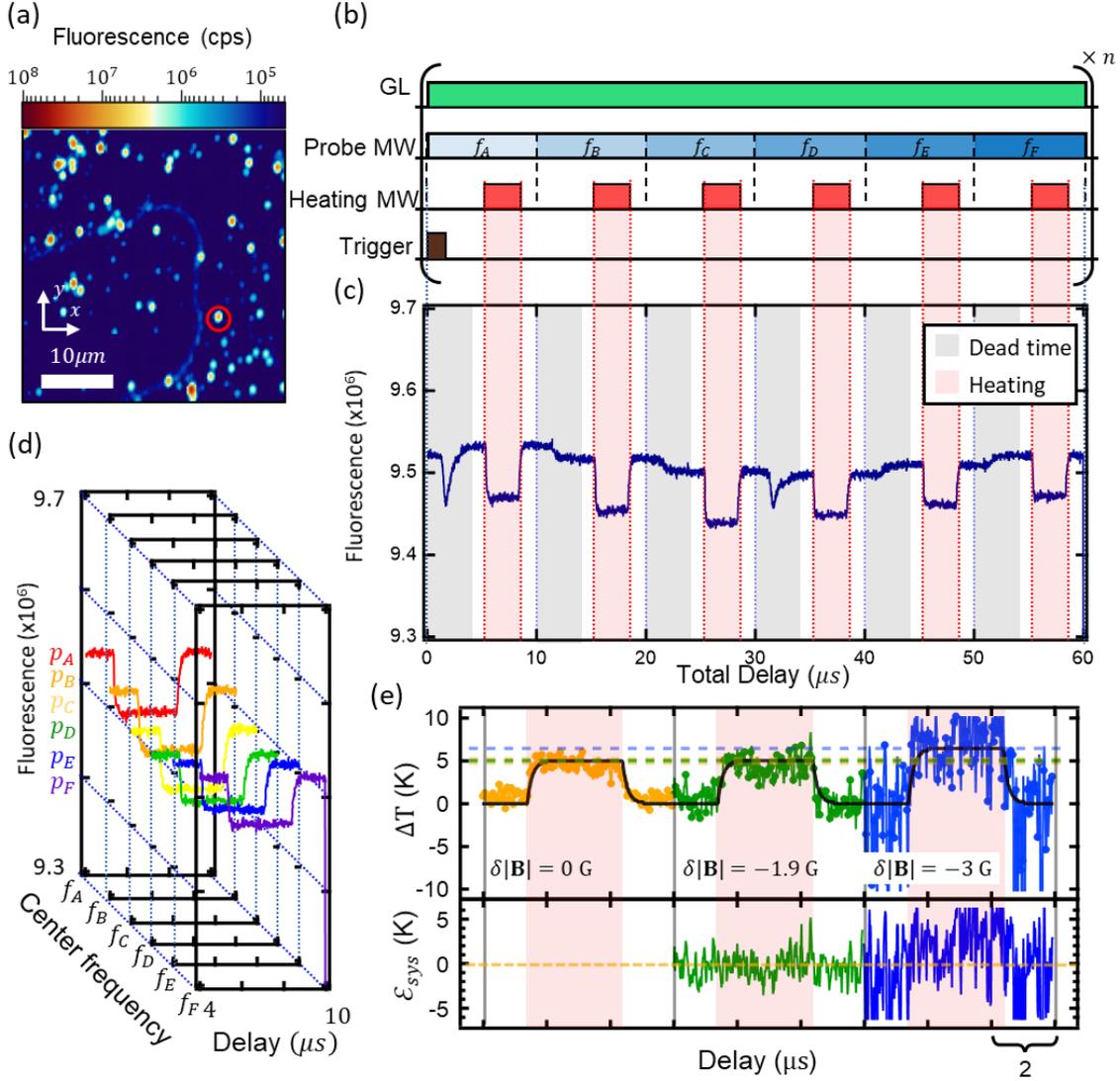

**Figure 3.** (a) Two-dimensional confocal scanning image of the NV centers in nanodiamonds dispersed on a cover glass with a shorted CPW fabricated on top. The temperature was measured at the marked NV center ensemble (red circle). (b) The microwave sequence used for the pump-probe temperature measurement. GL represents the green laser, Probe MW represents the change of the center frequency of the two microwaves used for DS-ODMR, Heating MW is the microwave pulse used for heating and $n$ is the number of accumulation sequence executed. (c) Raw fluorescence data measured using the time-tagged photon counting module. Dead time is the time range within the rise time of the frequency shifting and Heating is the time range where the heating microwave is on. (d) Fluorescence data

stacked based on delay. The fluorescence data sharing the same delay are grouped and used when converted into temperature. (e) Time-resolved measurement results using the six-point DS-ODMR method under different levels of external magnetic field detuning $\delta|\boldsymbol{B}|$. Each color of the dashed lines represents a temperature change value at the end of the heating MW pulse. The overlaid black line represents the simulation results for the converted temperature when the six-point measurement method is applied to the temperature changes calculated by HFSS and Python (see Supporting Information S3 and S5 for details). Upper: The measured temperature. Lower: The error induced by the magnetic field detuning, calculated by subtracting the $\delta|\boldsymbol{B}| = 0$ temperature curve from each of the other upper curves. ΔT represents the measured change in temperature and $\varepsilon_{sys}$ represents the systematic error.

The expected distortion to the measured temperature when an external magnetic field fluctuation is present was systematically demonstrated by applying magnetic field detuning to the system. Figure 3e presents a comparison of the resultant time-resolved thermometry for different values of $\delta|\boldsymbol{B}|$. The results measured with no magnetic field detuning reveal a root-mean-squared (RMS) temperature fluctuation of approximately 0.6 K. Combining this result with the integration time for each data point of 43 s(number of accumulation sequence $n$ = 1.5 × $10^8$) leads to a sensitivity of 3.7 $K \cdot Hz^{-1/2}$, which is above the shot-noise-limited sensitivity of 2.5 $K \cdot Hz^{-1/2}$ expected for this NV center ensemble. Therefore, a temperature change of approximately 5 K initiated by a pulsed microwave was measured with a signal-to-noise ratio (SNR) of approximately 8.3. Additionally, a rise time of approximately 300 ns can be resolved by our continuous-measurement-based time-resolved thermometry, where for the present setup, the time resolution is set by a 48 ns time step. As shown in the middle panel of Figure 3c, the result is robust to $\delta|\boldsymbol{B}|$ up to approximately 2 G, demonstrating a systematic error of 0.1 K in the measured temperature change due to $\delta|\boldsymbol{B}|$ which is less than the RMS temperature fluctuation, which is in agreement with the QuTip simulation results of the six-point measurement method applied to the temperature simulation results represented by a black solid line. (see below for simulation details, see also Supporting Information S3 and S5.) Therefore, for an external magnetic field fluctuation within the amplitude of 2 G, we expect the correspondingly induced measured temperature fluctuation to be within 0.1 K. This result is in agreement with the robust magnetic field range presented in Figure 2. The increase in the temperature fluctuation observed in the $\delta|\mathbf{B}|$ = -1.9 G result comes from the reduced sensitivity as the NV ensemble in the 100 nm nanodiamond slowly became unstable while the experiment proceeded. When $\delta|\mathbf{B}|$ is approximately 3 G, the systematic error is significant, reflecting the deviation demonstrated in the dressed-state picture as discussed above. Additionally, the SNR is less than 3 in this regime because the six-point measurement method becomes unstable when $\delta_B \sim \frac{\Gamma}{\sqrt{3}}$. The 2 G window is large enough for the measurement method to be robust to magnetic field fluctuations that may exist in practical measurement setups, such as fluctuations of the magnetic field strength of the neodymium magnet used for external magnetic field generation where approximately 0.1 G fluctuation is generated for a 1 K ambient temperature fluctuation,[49] or such as the fluctuations of the magnetic field generated from the currents of nanoelectric devices where a field of approximately 0.5 G is generated when a current of $I$ = 0.8 mA is injected into graphene.[42]

We will now discuss spatiotemporally resolved thermometry. Under the same pulsed heating scheme, **Figure 4**a presents three different NV centers distributed over the CPW, where the transient temperature increases over time differ, which can be inferred from the numerical simulations of the temperature distribution presented in Figure 4b. For the time-resolved measurement results, $n = 6 \times 10^8$ was used for ND 1 whose DS-ODMR curve showed a lower sensitivity of 9.4 $\mathrm{K \cdot Hz^{-1/2}}$, and $n = 1.5 \times 10^8$ was used for both ND 2 and 3 whose DS-ODMR curve showed a sensitivity of 3.6 $\mathrm{K \cdot Hz^{-1/2}}$. The numerical simulations were performed using combinations of microwave loss calculations based on HFSS software and heat conduction calculations based on Fourier's law along the surface of a physical environment with a convection term using Python (see Supporting Information S5 for details). Figure 4c presents an overlay of the simulation results (solid and dashed curves) and experimental results (symbols) for the temperature changes during microwave heating. By applying the convection term $Q_{\text{convection}} = h_{\text{convection}}(T - T_\infty)$, where $h_{\text{convection}}$ is the convective heat transfer and $T_\infty$ is the temperature at the boundary, 26 ºC for this experiment, the experimental result exhibits good agreement with $h_{\text{convection}} = 2 \times 10^8$ $\mathrm{W \cdot m^{-2} \cdot K^{-1}}$. The results demonstrate that this method can be applied to various orientations of NV ensembles in a common setting, particularly under a common external magnetic field, which is made possible by the properties of the dressed state discussed above.

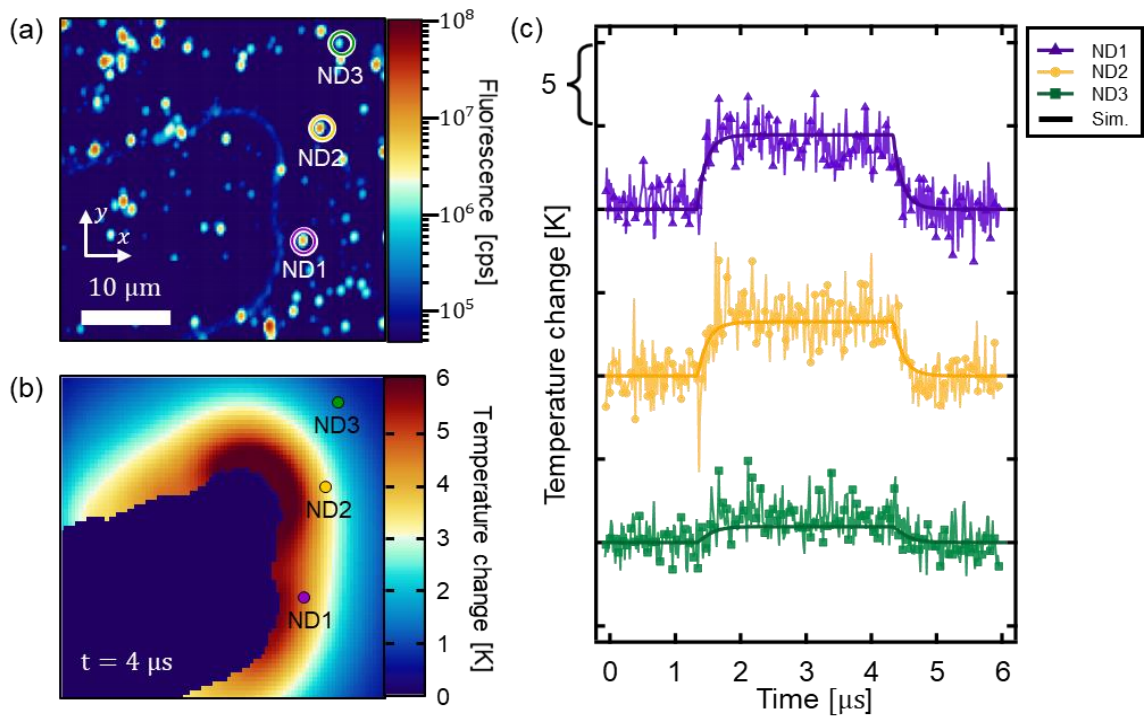

**Figure 4.** (a) Two-dimensional confocal scanning image of the NV centers in nanodiamonds dispersed on a gold CPW. The temperature was measured at the three marked NV center ensembles (colored circles). (b) Simulation results for the temperature distribution of the system when the temperature is saturated while the heating pulse is on. The colored dots represent the locations of the three nanodiamonds used for temperature measurement. (c) Overlaid experimental and simulation results for the time-resolved temperatures at the three measured NV centers.

In this work, we only evaluated the spatial resolution using nanodiamonds at a distance of approximately 10 μm apart from each other. In principle, the spatial resolution can be reduced to the diffraction limit of a confocal microscopy system, which could be implemented by increasing the density of nanodiamonds on the CPW surface. When applying this method, a pre-selection process for NV ensembles is required because the resonance peaks from different axes may overlap, making it difficult to apply the DS-ODMR to a single axis independently. This limitation can be overcome by attaching a single nanodiamond particle to a probe and performing scanning for 2D temperature imaging. Even with these limitations, we measured the spatially resolved transient temperature of the CPW surface during the heating and cooling process on the gold surface with high thermal conductivity. This result shows that the transient temperature imaging induced from heat generated by electronic devices, even ones with high thermal conductivity, may be measured using this method.

## 3. Conclusion

We developed a novel thermometry method based on DS-ODMR measurements and pump-probe measurements, which provide high spatiotemporal resolution while maintaining robustness to external magnetic field environments. By using a dressed spin state basis for thermometry, we obtained a broad range of initial external magnetic fields that exhibited attenuated reactions to magnetic field fluctuations within the range satisfying $\delta_B \ll \sqrt{2}\Omega$. Additionally, the applicability of this basis to the cw-ODMR-type measurement scheme enables temperature measurement with high temporal resolution. By combining the DS-ODMR method with a pump-probe temperature measurement sequence and by adopting the six-point measurement method with time-tagged measurements for a reasonable total measurement time, we were able to resolve the transient temperature inducted by a microwave pulse applied to a gold CPW with a time resolution of approximately 50 ns and temperature precision of 0.6 K. The insensitivity of the dressed spin state to external magnetic fields facilitated the application of this method to arbitrarily dispersed nanodiamond particles and allowed us to demonstrate spatiotemporally resolved temperature measurement using several nanodiamonds on the CPW surface. Thus, our thermometry method can be applied to the measurement of time-resolved temperature using nanodiamonds in complex environments such as living cells, where aligning an external magnetic field to the NV center axis of each independent nanoparticle is difficult. In addition, because a continuous microwave is used, it is easy to measure the sample temperature as a function of microwave power before the experiment, and one can avoid damage to the biological tissue by setting the initial temperature of an incubator considering microwave heating.[50] The method is also expected to facilitate investigating the thermal properties of nanoelectronic devices where external magnetic fields fluctuate over time by the varying electric current. Therefore, the high temporal resolution and selective temperature sens itivity demonstrated here can be applied to the characterization of the transient thermal processes occurring in various nanoscale systems.

**Supporting Information**

Supporting Information is available from the Wiley Online Library or from the author.

**Acknowledgements**

This work was supported by the National Research Foundation of Korea (NRF) Grant funded by the Korean Government (MSIT) (No. 2018R1A2A3075438, No.2019M3E4A1080144, No.2019M3E4A1080145, and No.2019R1A5A1027055) and the Creative-Pioneering Researchers Program through Seoul National University (SNU). Jiwon Yun and Kiho Kim contributed equally to this work.

# Supporting Information

# Temperature selective thermometry with sub-microsecond time resolution using dressed-spin states in diamond

*Jiwon Yun‡, Kiho Kim‡, Sungjoon Park and Dohun Kim\**

*Department of Physics and Astronomy, and Institute of Applied Physics, Seoul National University, Seoul 08826, Korea*

‡ These authors contributed equally to this work.
\*Corresponding author: <dohunkim@snu.ac.kr>

## 1. Theoretical analysis on the dressed-spin states

The Hamiltonian of the NV center under arbitrary external magnetic field is given by,

$$H_0 = D(T)S_z^2 + \gamma_e \boldsymbol{B} \cdot \boldsymbol{S} + E(S_x^2 - S_y^2) \qquad (S1)$$

where $D(T)$ is the zero-field splitting of the NV center, $\gamma_e$ is the gyromagnetic ratio of the electron spin, $\boldsymbol{B}$ is the external magnetic field, and $\boldsymbol{S}$ is the vector form of spin-1 matrices, and $E$ is the off-axial strain. In matrix form,

$$H_0 = \begin{pmatrix} D(T) + \gamma_e B_z & \gamma_e B_x & E \\ \gamma_e B_x & 0 & \gamma_e B_x \\ E & \gamma_e B_x & D(T) - \gamma_e B_z \end{pmatrix} \qquad (S2)$$

where $B_z$ is the magnetic field component parallel to the NV center axis and $B_x$ is the perpendicular component. For the intermediate $|\boldsymbol{B}|$ range of $E \ll \gamma_e|\boldsymbol{B}|$ and $\gamma_e B_x \ll D(T)$, since the effect of $E$ is negligible compared to combined energy scale set by the zero field and external magnetic field terms in the Hamiltonian (see Figure S1), the Hamiltonian can be approximated as

$$H_0 = \begin{pmatrix} D(T)+\gamma_e B_z & \gamma_e B_x & 0 \\ \gamma_e B_x & 0 & \gamma_e B_x \\ 0 & \gamma_e B_x & D(T)-\gamma_e B_z \end{pmatrix} \quad (S3)$$

and when this Hamiltonian is diagonalized, it can be approximated as

$$H_{diag} \sim \begin{pmatrix} D(T)+\gamma_e B_z + \frac{1}{2}\frac{\gamma_e B_x}{D(T)}\gamma_e B_x & 0 & 0 \\ 0 & -\frac{\gamma_e B_x}{D(T)}\gamma_e B_x & 0 \\ 0 & 0 & D(T)-\gamma_e B_z + \frac{1}{2}\frac{\gamma_e B_x}{D(T)}\gamma_e B_x \end{pmatrix}$$

(S4)

up to the order of $O((\gamma_e |\boldsymbol{B}|/D)^3)$. We set the eigenbasis of the diagonalized Hamiltonian as $|+1\rangle$, $|0\rangle$, and $|-1\rangle$.

With external microwaves $H_{MW1} = \boldsymbol{\Omega}_1 \cdot \boldsymbol{S} \cos(\omega_1 t)$, $H_{MW2} = \boldsymbol{\Omega}_2 \cdot \boldsymbol{S} \cos(\omega_2 t)$, the total Hamiltonian in the doubly rotating frame becomes,

$$H_{tot} = \begin{pmatrix} \lambda_{(+)} - \omega_2 & \Omega_{1,xy} & 0 \\ \Omega_{1,xy} & 0 & \Omega_{2,xy} \\ 0 & \Omega_{2,xy} & \lambda_{(-)} - \omega_1 \end{pmatrix} \quad (S5)$$

where $\lambda_{(\pm)} \sim D(T) \pm \gamma_e B_z + \frac{3}{2}\frac{\gamma_e B_x}{D(T)}\gamma_e B_x + O\left(\left(\frac{\gamma_e |\boldsymbol{B}|}{D(T)}\right)^3\right)$ is the energy level splitting in each of the subspaces, $\omega_1$, $\omega_2$ is the microwave frequency applied to the $|0\rangle$ - $|-1\rangle$ subspace and $|0\rangle$ - $|+1\rangle$ subspace with the resonant condition is $\omega_1 = \lambda_{(-)}$ and $\omega_2 = \lambda_{(+)}$, and $\Omega_{1,xy}$ ($\Omega_{2,xy}$) is the amplitude perpendicular to the quantization axis of each subspace of the microwave with frequency $\omega_1(\omega_2)$. For the temperature measurement method in this study, we tune the perpendicular amplitude so that $\Omega_{1,xy} = \Omega_{2,xy} = \Omega$.

With detuning in $D(T)$ and $\boldsymbol{B}$ corresponding to changes in environmental factors, the Hamiltonian becomes,

$$H_{\text{tot}} = \begin{pmatrix} \delta D(T) + \gamma_e \delta B_z + 3\dfrac{\gamma_e B_x}{D(T)}\gamma_e \delta B_x & \Omega & 0 \\ \Omega & 0 & \Omega \\ 0 & \Omega & \delta D(T) - \gamma_e \delta B_z + 3\dfrac{\gamma_e B_x}{D(T)}\gamma_e \delta B_x \end{pmatrix} \quad (S6)$$

Moving to the dressed spin state basis, the Hamiltonian becomes

$$H_{\text{tot,d}} = \begin{pmatrix} \delta D(T) + 3\dfrac{\gamma_e B_x}{D(T)}\gamma_e \delta B_x & \sqrt{2}\Omega & \gamma_e \delta B_z \\ \sqrt{2}\Omega & 0 & 0 \\ \gamma_e \delta B_z & 0 & \delta D(T) + 3\dfrac{\gamma_e B_x}{D(T)}\gamma_e \delta B_x \end{pmatrix} \quad (S7)$$

which is the result used in the main text.

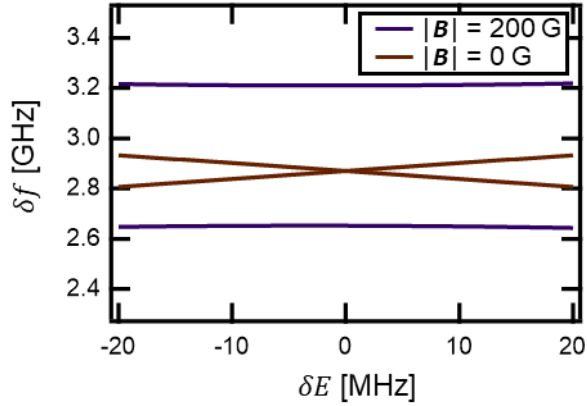

**Figure S1. Energy level difference of the spin transitions under change of *E*.** The energy level splitting becomes flat when an external magnetic field on the order of 100 G is present.

## 2. Experimental setup

We carried out the experiments in a homebuilt confocal microscope as schematically shown in Figure S2a. A green laser (CNI, MLL-III-532-200mW) is used for the measurement of the pulsed spin resonance characteristics of NV centers. The fluorescence emitted from NV centers is detected by an avalanche photodiode (APD, Excelitas, SPCM-AQRH-14-FC) which is connected to a two-channel gated photon counter (Stanford Research Systems, SR400). The external magnetic field was applied by locating a neodymium magnet fixed to a 3-axis stage around the diamond sample. The magnitude of the external magnetic field was varied by moving the 3-axis stage. To manipulate the electron spin state in the dressed frame, two driving microwaves are generated by signal generators (Stanford Research Systems,

SG396 and SG394), which are gated by two arbitrary function generators (AFGs) (Tektronix, AFG3252) and applied to the NV centers simultaneously. The microwave pulses amplified by a high-power microwave amplifier (Mini-Circuits, ZHL-16W-43+) are fed to a coplanar waveguide (CPW) fabricated on a coverslip. We prepared a type-Ib bulk diamond (Element 6) for external magnetic field reaction analysis using a single NV center and nanodiamonds with an average diameter of 100 nm (Adamas) dispersed on top of CPW for spatiotemporally resolved temperature-selective thermometry measurements.

In the time-resolved measurement, both the green laser and the microwaves from the two signal generators are applied continuously as shown in Figure S2b.
The output photons from the NV center are measured continuously using a time-tagged photon counting module (PicoQuant, TimeHarp 260 NANO). To measure the fluorescence at the 6 microwave frequencies of the Lorentzian curve, the analog frequency modulation mode is used in both signal generators and the input analog waveform was generated using the arbitrary waveform generator. The frequency modulation deviation we used was 16 MHz and the frequency modulation bandwidth was 100 kHz. Under this setting, we calibrated the stepwise voltage output from the AFG where $\pm 0.5$ V corresponds to $\pm 16$ MHz frequency detuning. Considering the frequency switching time which is about 3.5 μs, the first 4 μs of data for each center frequency period are discarded.

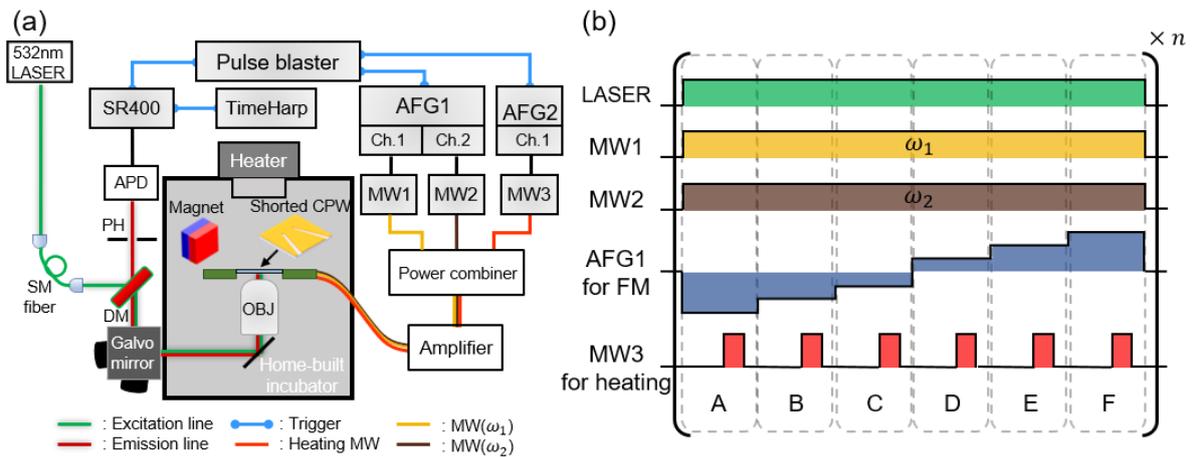

**Figure S2. Experimental setup (a)** Schematic setup for confocal microscopy and microwave setup. **(b)** Laser and microwave pulse sequence used for time-resolved measurement. APD: Avalanche photo diode, PH: pinhole, DM: Dichroic mirror, AFG: arbitrary function generator, MW: microwave signal generator, MW1 and MW2 are used for the dressed state ODMR measurement, and MW3 is used for heating. CPW: coplanar waveguide, OBJ: objective.

### 3. QuTip simulation of the dressed state cw-ODMR results

Quantum dynamics of the dressed spin-states were simulated using the QuTip package in written python. The fluorescence of the ODMR measurement sequence was simulated using the '*mesolve*' function in QuTip, which solves the master equation,

$$\frac{d\rho}{dt} = -i[H_{tot,d}, \rho] + \sum (L\rho L^\dagger - \frac{1}{2}\{L^\dagger L, \rho\}) \tag{S8}$$

with the Hamiltonian of Equation S7, where $\rho$ is the system density operator. We used the Lindblad operator

$$L_{-1}^{gl} = \sqrt{\Gamma_{gl}}|0\rangle\langle -1|, \ L_{+1}^{gl} = \sqrt{\Gamma_{gl}}|0\rangle\langle +1| \tag{S9}$$

where $\Gamma_{gl}$ is the decay rate coming from the green laser to reproduce the effect of the laser in the cw-ODMR measurement sequence. To simulate realistic cw-ODMR results, the master equation was calculated long enough for the quantum state to reach equilibrium ($\tau_{ODMR} > 1/\Gamma_{gl}$). We used $\Gamma_{gl} \sim 10$ MHz and the time range we used to solve the master equation to obtain the ODMR fluorescence was $\tau_{ODMR} \sim 10$ μs to match the experimental pulse repetition period.

### 4. The 6-point method for dressed state thermometry

By characterizing the Lorentzian resonance curve $L(f)$ of the dressed state cw-ODMR as a function of applied microwave frequency $f$, we convert the change in the photoluminescence (PL) of the six center frequencies $f_{A(B,C,D,E,F)}$ of the two applied microwaves into temperature difference. We choose these frequencies as follows.

$$f_A = (\omega_1 + \omega_2)/2 - \Gamma/2\sqrt{3} - d\omega,$$

$$f_B = (\omega_1 + \omega_2)/2 - \Gamma/2\sqrt{3},$$

$$f_C = (\omega_1 + \omega_2)/2 - \Gamma/2\sqrt{3} + d\omega, \tag{S10}$$

$$f_D = (\omega_1 + \omega_2)/2 + \Gamma/2\sqrt{3} - d\omega,$$

$$f_E = (\omega_1 + \omega_2)/2 + \Gamma/2\sqrt{3},$$

$$f_F = (\omega_1 + \omega_2)/2 + \Gamma/2\sqrt{3} + d\omega,$$

where $\Gamma$ is the linewidth of $L(f)$ and $d\omega$ is a detuning given to obtain the slope of the Lorentzian. The PL at each frequency can be approximated as,

$$p_A = p_{B,0} + \left.\frac{dL}{df}\right|_{f=f_B} \left(\frac{dD}{dT}\delta T - d\omega\right)$$

$$p_B = p_{B,0} + \left.\frac{dL}{df}\right|_{f=f_B} \left(\frac{dD}{dT}\delta T\right)$$

$$p_C = p_{B,0} + \left.\frac{dL}{df}\right|_{f=f_B} \left(\frac{dD}{dT}\delta T + d\omega\right) \quad (S11)$$

$$p_D = p_{B,0} - \left.\frac{dL}{df}\right|_{f=f_B} \left(\frac{dD}{dT}\delta T - d\omega\right)$$

$$p_E = p_{B,0} - \left.\frac{dL}{df}\right|_{f=f_B} \left(\frac{dD}{dT}\delta T\right)$$

$$p_F = p_{B,0} - \left.\frac{dL}{df}\right|_{f=f_B} \left(\frac{dD}{dT}\delta T + d\omega\right)$$

where $p_{A(B,C,D,E,F)}$ is the measured PL using $f_{A(B,C,D,E,F)}$ (see Figure S3), $p_{B,0}$ is the PL at the initial temperature at $f_B$, $dD/dT \sim 74$ kHz/K near 300 K, $\delta T$ is the change in the temperature, and $d\omega$ is the fixed detuning we used for preparing distinguished 6 center frequencies. Solving these equations, we obtain temperature conversion formula as,

$$\delta T = \frac{p_B - p_E}{(p_C - p_A)/2 - (p_F - p_D)/2} \frac{d\omega}{dD/dT} \quad (S12)$$

In addition, the shot-noise limited sensitivity of this method can be calculated from the slope of the Lorentzian curve as

$$\eta_T = 0.77 \frac{1}{dD/dT} \frac{\Gamma}{C_0} \frac{1}{\sqrt{R}} \tag{S13}$$

where $C_0$ is the contrast of the dressed state cw-ODMR, and $R$ is the measured photon count per second.

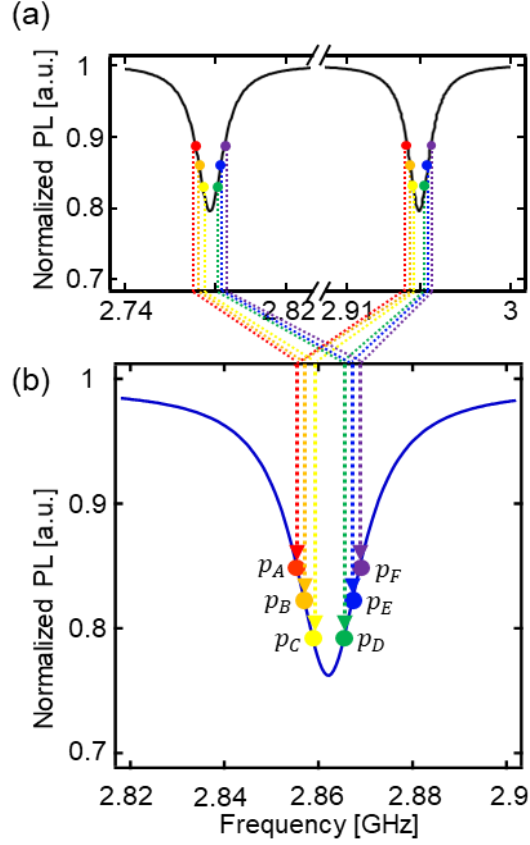

**Figure S3. (a)** Microwave frequencies applied to the NV center in the 6 point measurement for the DS-ODMR. **(b)** The corresponding points of the center frequencies represented in the DS-ODMR curve.

## 5. Temperature simulations

**Dielectric loss calculation**

To confirm the measured time-resolved temperature measurement, we conducted a numerical simulation of the temperature change. Based on the optical microscope image of the CPW, we replicated the geometrical structure. To describe the temperature change at the shorted-end of the CPW, the power losses attributed to the conductor and dielectric losses can be considered. We first calculated the power loss generated from the tapered and shorted coplanar waveguide design using the High-Frequency Structure Simulator (HFSS, ANSYS software), a simulator based on the finite element method, and obtained the surface loss density on the gold surface. The volume loss density of the substrate is neglected in this work as we are dealing with materials of low loss-tangent.

**Numerical heat transfer simulation**

Based on the calculated heat generation, we ran a numerical simulation using python. We designed a program to solve the 2-dimension partial differential equation with the Neumann boundary condition based on Fourier's law using discrete grids. The main equation used is

$$T(x,y,t+dt) = T(x,y,t) - \frac{dt}{\rho(x,y)C_{sp}(x,y)}(
\sigma(x,y)\frac{(T(x,y,t)-T_{-x}(x,y,t))}{(dx_-/2)((dx_++dx_-)/2)} + \sigma(x,y)\frac{(T(x,y,t)-T_{+x}(x,y,t))}{(dx_+/2)((dx_++dx_-)/2)}
+\sigma(x,y)\frac{(T(x,y,t)-T_{-y}(x,y,t))}{(dy_-/2)((dy_++dy_-)/2)} + \sigma(x,y)\frac{(T(x,y,t)-T_{+y}(x,y,t))}{(dy_+/2)((dy_++dy_-)/2)}
-Q_{mw}(x,y,t) + h_{convection}(T(x,y,t)-T_\infty)/l
) \qquad (S14)$$

where $\sigma(x,y)$ is thermal conductivity ($\sigma(x,y)=310$ W/m·K for gold, $\sigma(x,y)=0.026$ W/m·K for air), $C_{sp}(x,y)$ is specific heat ($C_{sp}(x,y)=0.129$ J/g·K for gold, $C_{sp}(x,y)=1.006$ J/g·K for air), $\rho(x,y)$ is the density ($\rho(x,y)=19.3\times 10^6$ g/m³ for gold, $\rho(x,y)=1.18\times 10^3$ g/m³ for air), $l = 500$ nm is thickness of the gold CPW we used and the thickness of the surface we consider to simplify the thermal system into 2D, $Q_{mw}(x,y,t)$ is the surface loss density calculated using HFSS, $h_{convection}$ is the convective heat transfer coefficient, $dx(y)_\pm$ is the size of the cell, $T_{(\pm x(y))}(x,y,t)$ is the temperature at the boundary of the grid containing point $x, y$ calculated as

$$T_{(\pm x)}(x,y,t) = \frac{\sigma(x,y)T(x,y,t)+\sigma(x\pm dx,y)T(x\pm dx,y,t)}{\sigma(x,y)+\sigma(x\pm dx,y)},$$

$$T_{(\pm y)}(x,y,t) = \frac{\sigma(x,y)T(x,y,t)+\sigma(x,y\pm dy)T(x,y\pm dy,t)}{\sigma(x,y)+\sigma(x,y\pm dy)}, \qquad (S15)$$

and the $\pm x(y)$ notations follow the grid scheme in Figure S4. We used the boundary condition as $T_\infty = 26°C$ to reflect the ambient condition.

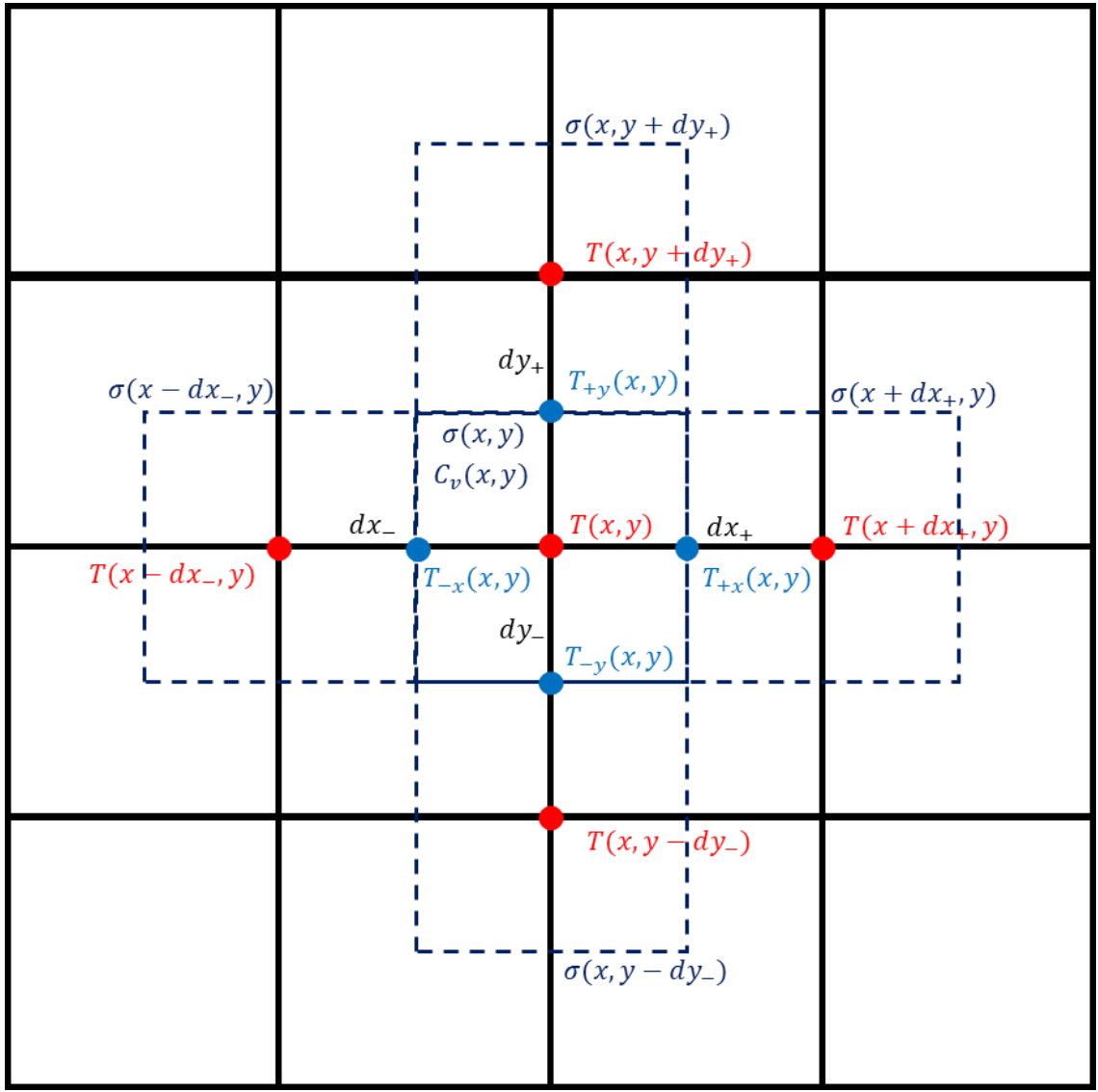

**Figure S4. Grid notation used for the temperature simulation using python.** The dotted line squares represent a single unit cell used in the temperature calculation.

28